\documentstyle[aps,pre,epsfig]{revtex}
\input epsf
\begin{document}

\title{The Scaling Behaviour of Stochastic Minimization Algorithms\\
in a Perfect Funnel Landscape}
\author{K. Hamacher and W. Wenzel}
\address{Institut f\"ur Physik, Universit\"at Dortmund, \\D-44221
Dortmund, Germany}
\maketitle
\begin{abstract} 
We determined scaling laws for the numerical effort to find the
optimal configurations of a simple model potential energy surface
(PES) with a perfect funnel structure that reflects key
characteristics of the protein interactions. Generalized Monte-Carlo
methods(MCM, STUN) avoid an enumerative search of the PES and thus
provide a natural resolution of the Levinthal paradox. We find that
the computational effort grows with approximately the eighth power of
the system size for MCM and STUN, while a genetic algorithm was found
to scale exponentially. The scaling behaviour of a derived lattice
model is also rationalized.
\end{abstract}


\vskip 0.3 in

Despite recent successes in the description of the molecular
structure\cite{m_pa1,gibson1} and the folding process of small
polypeptides\cite{vangunsteren,hansmann4} the {\em ab-initio}
prediction of the molecular structure for larger proteins remains an
elusive goal. Since sequencing techniques presently outperform
available experimental techniques for protein structure prediction
(PSP) by a wide margin, the reservoir of sequenced proteins of unknown
structure represents an ever growing pool of available, but as of yet
inaccessible, biological information.  These observations motivate the
search for {\em ab-initio} techniques to predict the molecular
structure of proteins from the amino acid sequence alone as one of the
outstanding challenges to biological physics.

In one widely pursued theoretical approach to PSP, the native
structure of the protein is sought as the global minimum of an
appropriate potential/free energy-function of the
molecule\cite{gibson1,scheraga5,skolnick1,derreumaux1} often including
interactions with the solvent in an approximate, implicit fashion. As
the folding process in nature takes place on a long time scale
($10^{-3} - 10$ s), its direct simulation cannot be accomplished with
the presently available computational resources. It is therefore
desirable to determine the global minimum of the potential function
without recourse to the folding dynamics. It has been argued that the
resulting minimization problem is
NP-hard~\cite{compcomplex,wille1,ngo1}, i.e.~that the number of
low-energy local minima grows exponentially with the number of amino
acid residues. For this reason stochastic minimization
procedures\cite{kirkpatrick1} are widely believed to be the most
promising avenue to avoid an exponential increase of the numerical
effort for the probabilistic ``solution'' to this problem. Since the
available computational resources fall short by orders of magnitude to
treat large proteins, it is important to obtain an order-of-magnitude
estimation of the numerical effort required. This question can be
answered by addressing the scaling laws\cite{gutin96,thirumalai3}:
\begin{equation}
        n_{\rm CPU} (N) \sim A \ N^\alpha,
\end{equation} 
governing the dependence of the computational effort ($n_{\rm CPU}$)
on the system size ($N$).

In this investigation we determined the scaling exponents for four
different global minimization methods, for a very simple, idealized a
model that reflects some key characteristics of the realistic
problem. Our results demonstrate that the Levinthal
paradox\cite{levinthal1,karplus1,shakhnovich1}, which arises from the
enourmous number of low-lying conformations of the protein, is
naturally resolved in the presence of a funnel structure. For such
models, stochastic, thermodynamically motivated, minimization
techniques are generically able to avoid the exponentially difficult
enumerative search of the potential energy surface (PES) in favour of
a power-law dependence. Our investigation of a novel stochastic
tunnelling technique, which removes the kinetic barriers between local
minima of the PES, demonstrates that the scaling exponents $\alpha$ is
determined by the thermodynamic complexity of the model, not by the
barrier height of the kinetic pathways. We find that the computational
effort of Monte-Carlo-based methods grows with approximately the
eighth power of the system size. The genetic algorithm we investigated
was the most efficient technique for small systems, but its
computational effort grew exponentially with system size. This finding
demonstrates that the investigation of the growth laws yields a much
stronger criterion for the selection of promising algorithms than the
comparison of different techniques for fixed system size. Finally, we
provide the first explicit demonstration that the scaling exponent of
Monte-Carlo techniques on a lattice model, which incorporates only the
low-energy physics of the continuum model, is consistent with its
continuum equivalent.

Because a detailed direct experimental characterization of the protein
PES is difficult, there is ample
controversy\cite{controversy,gutin96,li98} regarding its structure and
defining features. However, in recent years an consensus regarding the
existence of a ``funnel structure'' has emerged as the most important
characteristic of the PES in the present paradigm for protein
folding\cite{funnel,dill,karplus6}. In such a structure the global
minimum can be reached via a multitude of pathways that traverse a
sequence of increasingly well-formed intermediates in the folding
process. This observation implies a positive correlation between the
``distance'' of a given local minimum from the native state to the
relative energy difference between the two minima. There is some
evidence to suggest the existence of different families of protein
models within this paradigm\cite{gutin96,li98} which may be
characterized with different scaling laws in their folding
time. However, since the origins of these differences are presently
not known they are difficutl to incorporate into a simple continuum
model that remain amenable to treatment with present-day computational
resources. In order to determine a lower bound on the computational
complexity, we therefore focus on the scaling laws governing the
relaxation in a ``perfect funnel'' landscape. Such a landscape is
characteristic of the family of ``fast folders'' in the lattice
models. In addition to the existence of a funnel-structure we demand
that the PES reflects two other characteristics of their realistic
counterparts: a near-solid packing density in the vincinity of the
global minimum and the existence of two energy scales that are derived
from the two relevant types of interactions in polypeptides. The
free-energy difference between low energy protein conformations is
small (10 kcal/mol), arising from hydrogen-bonding, dispersion and
solvent interactions. In contrast, the energy barriers separating such
conformations are characterized by strong interactions ($\gg$ 100
kcal/mol), arising from covalent bonding and steric
repulsion. Simplifying significantly, the strong interactions are
responsible for the reduction of the phase space to a few
energetically allowed ``islands'', which are then differentiated in
energy by the weaker interactions.

{\em Model:} To obtain statistically relevant results for sufficiently
large systems, we have investigated a very simple two-dimensional
model, consisting of two types of particles that interact pairwise
with Lennard-Jones potentials of unit radius such that like particles
attract twice as strongly as unlike particles. The local minima of the
model PES are slight distortions of a triangular lattice. There are
exponentially many such minima, which are differentiated by the small
energy difference in the interaction strength of the two types of
bonds, while the transition states between the local minima are
characterized by the large energy scale of steric repulsion. The
problem is easily shown to be NP-hard\cite{wille1}. The dynamical
process by which a random initial condition develops to the minimal
configuration can be visualized as a ``demixing'' of the particles
into two adjacent clusters of particles of the same type --- the ideal
funnel structure of the global PES is thus obvious.  The average
distance any given particle must travel from a random initial
condition to its position in the minimal cluster grows with the system
size, mirroring the ``global'' transformations required to fold the
protein from the coiled to the native state.

We stress that the similarity between this model and the PSP is purely
abstract, there is no correspondence or mapping between the
coordinates of the particles and the coordinates of atoms or clusters
of atoms in the protein. Given that a ``global'' transformation is
required this minimization problem is more difficult than the
minimization of Lennard-Jones clusters studied previously\cite{m_pa1},
but lacks the specific one-dimensional constraints of various simple
protein models that have recently been studied on the
lattice\cite{PH-model}.  A lattice version of the model is easily
derived by associating each local minimum with its closest lattice
configuration.

{\em Methods:} As the basic technique we have investigated Monte-Carlo
with minimization (MCM)\cite{scheraga6,MCMSCHERAGA}, a generic and
parameter-free extension of simulated annealing
\cite{kirkpatrick1}, which accelerates the minimization-process by
allowing the configurations to relax locally before the Metropolis
criterion is applied. Since only the energies of the local minima are
compared to one another the simulation can proceed at a sufficiently
low temperature to differentiate the local minima. Our results for
trial runs using straightforward Monte-Carlo and simulated annealing
calculations and their recent generalizations\cite{multican,partemp}
showed that it would be impossible to obtain sufficiently good
statistics for $N_{\rm CPU}$ for large systems to estimate the scaling
behaviour.  

Secondly we investigated a novel stochastic tunnelling
method(STUN)\cite{levy1,barhen1}, where a transformed PES:
\begin{equation}
\tilde E(\vec{x}) = 1 - e^{-\gamma [E(\vec{x}) - E(\vec{x}_0)]},
\label{eq:tun}
\end{equation} 
is used in the dynamical process (Fig. 1). Since this transformation
compresses the energy interval above the currently optimal energy
$E(\vec{x}_0)$ into the interval [0,1], the high-energy scale of the
problem is effectively eliminated and the simulation self-adjusts its
``effective temperature'' as better and better configurations are
found.

Thirdly we have investigated a genetic algorithm (GA)~\cite{if_go1} as
a radically different approach to \nobreak stochastic global
minimization. From a population of size $P$, we select $P/2$ pairs of
configurations, each with probability
\begin{equation}
p_i = \frac{E_{\rm max} - E_i }{\sum_j (E_{\rm max} - E_j)}, 
\end{equation}
where $E_i$ designates the energy of configuration $i$ and $E_{max}$
the maximal energy of the present population. Two new configurations
are generated from each pair created by randomly exchanging
consecutive subsets of coordinates between the two configurations
(crossover). 

In addition a random alteration of one coordinate is  made
with a small probability (mutation). The latter step insures the
ergodicity of the method, but most novel configurations are generated
by the crossover mechanism. As a reference we have gathered data for
the multi-start algorithm (MS), where a sequence of independent random
initial conditions is subject to local minimization.

{\em Results:} As an unbiased measure of the efficiency of a
particular algorithm $N_{\rm CPU}$ we adopted the average number of
function evaluations ($n_{90}$) that is required to locate the global
minimum with 90\% probability. Given a set of parameters, we conducted
between 100 and 500 independent runs. We heuristically determined a
run-size $n_{\rm max}(N)$ for which well over 90\% of the runs were
able to locate the global minimum. From this data we directly
determined the fraction of runs necessary to locate the minimum
$n_{\rm 90,raw}$. Because of the (asymptotic) time invariance of the
minimization algorithms, the first-passage probability $p(n)$ must
obey an exponential distribution. The systematic error to $n_{\rm 90,
raw}$ is therefore small, the details of the data analysis will be
published elsewhere\cite{long-paper}. For even system sizes $N=4 - 16$
we have optimized the parameters of the various methods. The optimal
parameters were found to be only slightly dependent on system size and
could be extrapolated to larger system sizes where a full parameter
optimization was too expensive\cite{long-paper}. 

For just under a decade of system sizes (Figure~\ref{fig:scaling}) we
obtain a power-law dependence of the computational effort with the
system size with scaling exponents as $\alpha_{\rm STUN} = 7.6 (\pm
1.8)$ and $\alpha_{\rm MCM} = 6.4 (\pm 1.5)$ for the continuum and
$\alpha_{\rm MC/MCM} = 4.7 (\pm 1.6)$ for the lattice model.
The slight curvature of the MCM data for large system size correlates
with an increasing efficiency of the local minimization algorithm we
used (inset of Figure~\ref{fig:scaling}). Taking into account the
exponents of the local minimization method, which scales almost
linearly in the range of system size investigated, we find
$\alpha_{\rm MC,lattice} \approx
\alpha_{\rm MCM} -\alpha_{\rm conj. gradient}$.  For the GA and MS an
exponential increase of the computational effort $n_{\rm 90,raw} \sim
e^{\xi N}$ with system size was observed., with exponents $\xi_{MS} =
0.64$ and $\xi_{GA} = 0.37$

{\em Conclusions:} The demonstration of power-law growth of the
computational effort for the Monte-Carlo method (MCM) illustrates the
fact that a the existence of a funnel structure is sufficient to avoid
an exponentially expensive search of the PES. This observation offers
a natural resolution of the Levinthal paradox in the context of
thermodynamically motivated, stochastic minimization methods: The
exponential complexity in the Levinthal paradox results from the
assumption that the local minima appear as uncorrelated ``holes'' on
an otherwise flat PES. Obviously, the enumerative search of such a PES
is unavoidable. The two necessary ingredients for a power-law scaling
of the ``folding time'' are the existence of a hierarchy of the local
minima and a method that can exploit this hierarchy by virtue of the
correlation of successive configurations.  The key difference between
MS and MCM lies in the lack of correlation between the configurations
of the former method and results in the expected exponential increase
of the numerical effort for MS.

The equivalence of the exponents of MCM and the tunnelling method,
which systematically eliminates kinetic barriers in the minimization
process, indicate that presence and height of such barriers do not
affect the scaling behaviour of the method. It is therefore the
thermodynamic complexity of the PES, as opposed to the presence of
kinetic constraints, which classifies the folding process here. This
observation raises the intriguing question, whether the scaling
exponents are different if the structure of the minima of the PES is
altered in the transformation, such as in the diffusion equation
method\cite{scheraga2}.

We note that the superiority of MCM over GA can only be established in
the context of a scaling analysis, as the GA is the superior method
for small system size. The reasons for the failure of the GA are
presently ill-understood.  Compounded with the $N^2$ effort to
evaluate a long-range pair-potential, the total minimization effort
grows with the eighth power of the system size, which places the
protein structure problem among the computationally hardest problems
studied today. In the context of recent discussion regarding the
``foldability''\cite{shakhnovich1,irbaeck1,gutin96} of different
families of model-``proteins'', our model is a natural ``fast-folder''
by virtue of construction.  It is therefore encouraging that our
results offer the first explicit confirmation that the scaling
behaviour of the continuum systems is consistent with the behaviour of
the derived lattice model, while the numerical effort of the treatment
of the latter is orders of magnitudes less. It is further encouraging
that the scaling exponent for the continuum model agrees within the
statistical error with estimations of the ``folding-time'' in polymer
models\cite{thirumalai3} and some lattice models for
proteins~\cite{gutin96}, provided that the number of local minima
visited in the first-passage trajectory is proportional to the folding
time. We hope that our observations motivate the investigation of
scaling laws for more realistic models and a wider variety of
methods, when the computational resources for such investigations
become available. The study of models that incorporate
the one-dimensional connectivity of protein molecule in the presence
of various types of interaction will allow to differentiate between
the various mechanisms that have been postulated to aide the folding
process in nature. Beyond the PSP problem, NP-hard minimization
problems are ubiquitous in many scientific and industrial
areas~\cite{compcomplex} and it would be highly desirable to establish
``universality classes'' for such problems, which are characterized by
their scaling exponent $\alpha$.

{\em Acknowledgements:} The authors gratefully acknowledge
stimulating discussions with H. Keiter, J. Stolze, C. Gros, M. Karplus
and U. Hansmann. This work was supported by the state government of
NRW.

\bibliographystyle{prsty}

\begin{figure}
\vspace{-1.5cm}
\centerline{\epsfig{figure=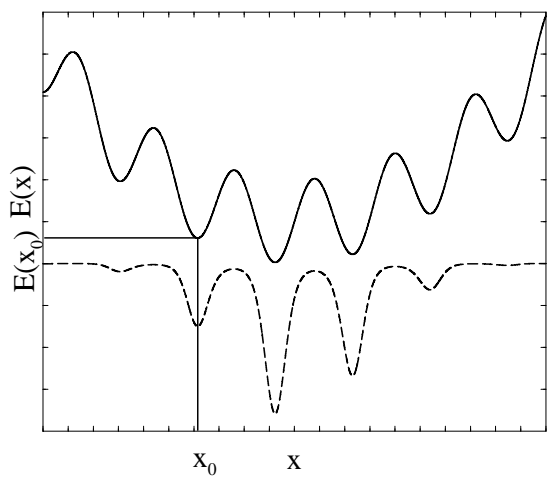,height=9cm,angle=-90}}
\caption{Schematic one-dimensional PES (full line) and its STUN effective
potential(dashed line), where the indicated minimum
$E(\vec{x}_0)$ is used as the reference. All energies ranging from the
best present estimate to infinity are mapped to the interval
$[0,1]$, while all the energies of all lower minima are exponentially
enhanced.
\label{fig:transform}}
\end{figure}

\begin{figure}[b]
\vspace*{-0.8cm}
\centerline{\epsfig{figure=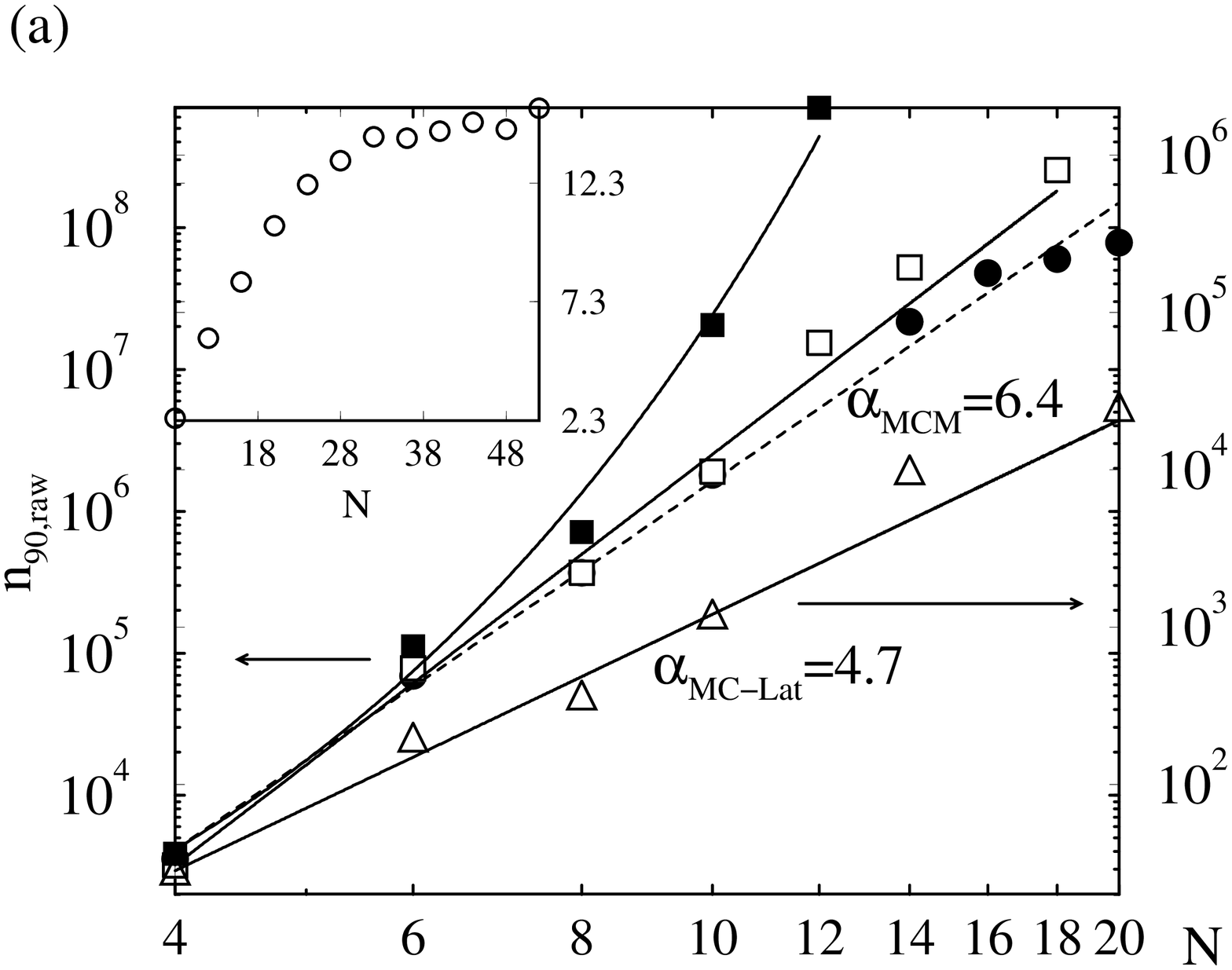,height=8cm,angle=-90}}
\vspace*{-1cm}
\centerline{\epsfig{figure=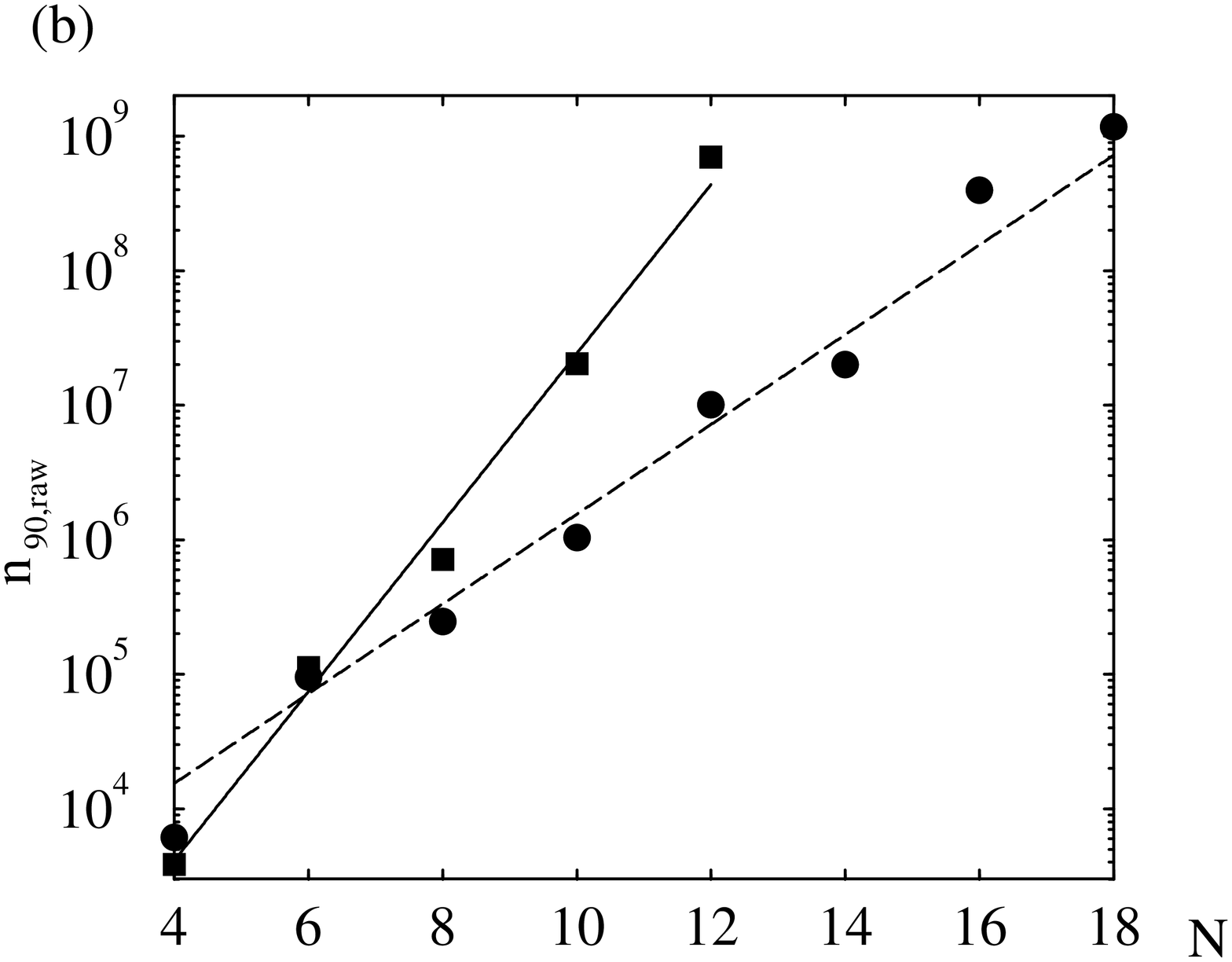,height=8cm,angle=-90}}
\vspace*{0.1cm}
\caption{(a) log-log plot of the average number of function
evaluations $n_{\rm 90,raw}$ as a function of system size N for
Monte-Carlo with minimization MCM (circles) and the stochastic
tunnelling method STUN (open squares) in the continuum (left scale)
and for Monte-Carlo(triangles) on the lattice (right scale) with
power-law fits. The inset shows the average number of
function evaluations (in thousands) for the minimization of
a cluster of N particles using the conjugate gradient algorithm. To
demonstrate that exponential and power-law scaling can be clearly
distingusiched we show data for the exponentially scaling MS
algorithm (full squares).  (b) log-linear plot of $n_{\rm 90,raw}$(N)
for the multi-start method (MS) (squares) and the genetic algorithm
(GA) (cirles) with exponential fits.
\label{fig:scaling}}
\end{figure}

\end{document}